\documentstyle[12pt]{article}

\begin{document}
\begin{titlepage}
\title{Sigma Models and Minimal Surfaces}
\vspace{5cm}
\author{Metin G{\" u}rses \\
{\small Department of Mathematics, Faculty of Science} \\
{\small Bilkent University, 06533 Ankara - Turkey}\\
email:gurses@fen.bilkent.edu.tr}
\maketitle
\begin{abstract}
The correspondance is established between the sigma models , the minimal
surfaces and the Monge-Amp{\' e}re equation. The Lax-Pairs of the minimality
condition of the minimal surfaces
and the Monge-Amp{\' e}re equations are given. Existance of
infinitely many nonlocal conservation laws is shown and some
Backlund transformations are also given.
\end{abstract}
\end{titlepage}

\noindent
{\bf 1.} \,In a recent paper \cite{KAR} , we have investigated the classical
integrability of the sigma models in a non-riemannian background
and have given their one soliton Backlund transformations. In
particular , two dimensional sigma-models with a Wess-Zumino term
have been studied in detail.

Let $M$ be a 2-dimensional manifold with local coordinates
$x^{\mu}=(t,x)$
and $\Lambda^{\mu \nu}$ be the components of a tensor field in $M$. Let P be
an $2 \times 2$ matrix with $det(P)=1$. We assume that $P$ is a
hermitian ($P^{\dagger}=P$) matrix. Then the sigma-model we
consider is given as follows

\begin{equation}
{\frac{\partial}{\partial x^{\alpha}}}\,\Bigl(\Lambda^{\alpha \beta} P^{-1}
{\frac{\partial\, P}{\partial x^{\beta}}}\Bigr) = 0. \label{SM}
\end{equation}

\noindent
The integrability of the above equation has been studied in \cite{KAR}.
The uniqueness of the solutions of these equations under certain boundary
conditions is given in \cite{GUR}. In these works
the matrix function $P$ and the tensor
$\Lambda^{\alpha\, \beta}$ were considered independent. We have classified
possible forms of the tensor $\Lambda^{\alpha \, \beta}$ under the condition
of integrability.

In some cases these two quantities may be related. Such a relation
may provide some intersting equations. In this work we are
interested in the integrability property of such cases.
As an example ,let $P=g$ where $g$ is matrix representing
the metric $g_{\alpha \, \beta}$ , symmetric with respect to
the lower indices. Letting also $\Lambda^{\alpha \, \beta}=
g^{\alpha \, \beta}$ , the inverse components of the metric
 $g_{\alpha \, \beta}$ , then (\ref{SM}) becomes

\begin{equation}
{\frac{\partial}{\partial x^{\alpha}}}\,\Bigl(g^{\alpha \beta} g^{-1}
{\frac{\partial\, g}{\partial x^{\beta}}}\Bigr) = 0. \label{SM1}
\end{equation}

In the theory of surfaces in $R^{3}$ there is a class , the minimal surfaces
which have special importance both in physics and mathematics \cite{BU1},
\cite{die}. Let $S=\{ (t,x,z) \varepsilon R^{3}; z=h(t,x)\}$ define a surface
$S \varepsilon R^3$ which is the graph of a differentiable function $h(t,x)$.
This surface is called minimal if $h$ satisfies the condition

\begin{equation}
(1+h,_{x}^2)\,h,_{tt}-2h,_{x}\,h,_{t}\,h,_{xt}+(1+h,_{t}^2)\,h,_{xx}=0,
\label{met5}
\end{equation}

\noindent
The Gaussian curvature $K$ of the surface $S$ is given by

\begin{equation}
K={h,_{xx}\, h,_{tt}-h,_{xt}^2 \over (1+h,_{x}^2+h,_{t}^2)^2}
\end{equation}

\noindent
{\bf 2.} \,The sigma model equation (\ref{SM}) is integrable for certain choices of
the tensor field $\Lambda^{\alpha \, \beta}$. In two dimensions the
integrability conditions on this tensor are given by

\begin{equation}
\partial_{\alpha}\,({1 \over \sigma}\, \Lambda^{\alpha\, \beta}\,
\partial_{\beta} \, \sigma)=0~~,~~
\partial_{\alpha}\,({1 \over \sigma}\, \Lambda^{\beta \,\alpha}\,
\partial_{\beta} \, \phi)=0.
\end{equation}

\noindent
where $\sigma$ is the determinant  and $\phi$ is its antisymmetric part
 of the tensor field  $\Lambda^{\alpha \, \beta}$.
Hence  by letting $\Lambda^{\alpha \, \beta}=g^{\alpha \, \beta}$
the above conditions are trivially satisfied becouse $\sigma=1$ and
$\phi=0$. Then using the approach developed in \cite{KAR} it is
straightforward to show that (\ref{SM1}) is also integrable. This leads to
the following proposition.

\noindent
{\bf Proposition 1}: The Lax pair of (\ref{SM1}) is

\begin{equation}
\epsilon^{\alpha\, \beta}\, \frac{\partial}{\partial x^{\beta}}\,\, \Psi=
{1 \over k^2+1}\,(k\,g^{\alpha\, \beta}-\epsilon^{\alpha\, \beta})\,g^{-1}\,
\frac{\partial\,g}{\partial x^{\beta}}\,\, \Psi \label{lp1}
\end{equation}

\noindent
provided $det(g)=1$ and $g_{\alpha\, \beta}$ is symmteric. Here $k$ is
an arbitrary constant (the spectral parameter),
$\epsilon^{\alpha \, \beta}$ is the Levi-Civita tensor with
$\epsilon^{12}=1$.

\noindent
A standard parametrization of $g_{\alpha \, \beta}$ may be given as follows

\begin{equation}
ds^2 = g_{\alpha\, \beta}\,d\,x^{\alpha}\, d\,x^{\beta}\\
={1 \over w}\,[(1+a^2)\,d\,t^2+2\,a\,b\,\,d\,x\,d\,t+(1+b^2)\,d\,x^2]
\label{met1}
\end{equation}

\noindent
where $x^\alpha=(t,x)$, $a$ and $b$ are differentiable functions of
$t$ and $x$ and

\begin{equation}
w^2=1+a^2+b^2. \label{ro}
\end{equation}

\noindent
{\bf Proposition 2:} Let $h$ be a differentiable function of $t$ and $x$
and let $a=h_{,t}$ and $b=h_{,x}$ , then the
minimality condition (\ref{met5}) solves the sigma model equation
(\ref{SM1}).

\noindent
This result is ineteresting and also very important. We shall give
the Lax-pair (\ref{lp1}) in a more detailed way,
but before that we write the minimality condition in a covariant way.
The metric on this minimal two dimensional surface $S$ is

\begin{eqnarray}
(ds)_{m}^2=g_{(m)\, \mu\, \nu}\,dx^{\mu}\,dx^{\nu}\\
=(1+h,_{t}^2)\,dt^2+2\,h,_{t}\,h,_{x}\,\,dx\,dt+
(1+h,_{x}^2)\,dx^2 \label{met6}
\end{eqnarray}

\noindent
Then the minimality condition (\ref{met5}) may be written covariantly as

\begin{equation}
g_{(m)}^{\alpha \, \beta}\,\partial_{\alpha}\, \partial_{\beta}\,h=0.
 \label{har1}
\end{equation}

\noindent
Since $g_{(n)\,\,\mu \nu}=\delta_{\mu \nu}+h,_{\mu}\, h,_{\nu}$ , where
$\delta_{\mu \nu}$ is the Kronecker delta symbol ,
(\ref{har1})  is also equivalent to

\begin{equation}
\partial_{\alpha}\,(\sqrt{g_{(m)}}\,g_{(m)}^{\alpha \, \beta})=0.
\label{har2}
\end{equation}

\noindent
where $g_{(m)}$ is the determinant of the metric $g_{(m)\,\alpha\, \beta}$ on $S$.
$S$ is embedded in a flat three dimensional Euclidean space $R^{3}$
with metric $ds^2=dt^2+dx^2+dz^2$. The minimality conditions (\ref{har1})
and(\ref{har2}) are equivalent to the harmonicity of the function $h(t,x)$
with respect to the metric of $S$

\begin{equation}
\partial_{\alpha}\,(\sqrt{g_{(m)}}\,g_{(m)}^{\alpha \, \beta}\,
\partial_{\beta}\,h)=0. \label{har3}
\end{equation}

\noindent
In the language of harmonic mappings of riemannian manifolds \cite{eel}
Eqns(\ref{har1}), (\ref{har2}) , and
(\ref{har3}) imply that the mapping $x^{\alpha}: S \rightarrow S$ is
harmonic.
Here we would like remark that the nonlinear partial differential equation
(\ref{met5}) describing the minimality condition of a two dimensional
surface $S$ is a special case of the sigma model equation (\ref{SM1}).
Hence it straightforward to conclude that the Eq.(\ref{met5}) is integrable
and its Lax-pair is given in (\ref{lp1}). We shall now give this
Lax-pair more explicitly. Let $A=g^{-1}\, \partial_{t}\, g$ and
$B=g^{-1}\, \partial_{x}\, g$ be two $2 \times 2$ matrices with components

\begin{eqnarray}
A^{1}_{1}={1 \over w^2}\,[p(1+q^{2})r-q(1+p^{2})s]\\
A^{1}_{2}={1 \over w^2}\,[q(1+q^{2})r+p(1-q^{2})s]\\
A^{2}_{1}={1 \over w^2}\,[q(1-p^{2})r+p(1+p^{2})s]\\
A^{2}_{2}=-{1 \over w^2}\,[p(1+q^{2})r-q(1+p^{2})s]
\end{eqnarray}

\begin{eqnarray}
B^{1}_{1}={1 \over w^2}\,[p(1+q^{2})s-q(1+p^{2})t]\\
B^{1}_{2}={1 \over w^2}\,[q(1+q^{2})s+p(1-q^{2})t]\\
B^{2}_{1}={1 \over w^2}\,[q(1-p^{2})s+p(1+p^{2})t]\\
B^{2}_{2}=-{1 \over w^2}\,[p(1+q^{2})s-q(1+p^{2})t]
\end{eqnarray}

\noindent
where we have used the same notation used in \cite{die}

\begin{eqnarray}
p=h_{t}~,~q=h_{x}~,~r=h_{tt}~,~s=h_{tx}~,~t=h_{xx}\\
w^2=1+p^2+q^2
\end{eqnarray}

Then the Lax-pair becomes

\begin{eqnarray}
\Psi_{,x}=-{1 \over k^{2}+1}\,[k(-r^{\prime}\,A+q^{\prime}\,B)+B]\, \Psi \label{l1}\\
\Psi_{,t}=-{1 \over k^{2}+1}\,[k(-q^{\prime}\,A+p^{\prime}\,B)+A]\, \Psi   \label{l2}
\end{eqnarray}

\noindent
where $k$ is the spectral parameter  $p^{\prime}$ , $q^{\prime}$ and
$r^{\prime}$ are given by

\begin{equation}
p^{\prime}={1+p^{2} \over w}~~,~~q^{\prime}={p\,q \over w}~~,~~
r^{\prime}={1+q^{2} \over w}
\end{equation}

\noindent
Integrability of the equations (\ref{l1}) and (\ref{l2}) give

\begin{eqnarray}
(r^{\prime}\,A-q^{\prime}\,B)_{,t}+(p^{\prime}\,B-q^{\prime}\,A)_{,x}=0\\
A_{,x}-B_{,t}=[A,B]
\end{eqnarray}

\noindent
The first of the above equation is identical with the minimality
condition (\ref{met5}) and the second one is a trivial identity.

\noindent
{\bf 3.} \,From the Lie symmetries of the minimality condition it may be possible
to find some conservation laws. Some of these are given by \cite{die}

\begin{eqnarray}
({q \over w})_{,x}+({p \over w})_{,t}=0\\
({p\,q \over w})_{,x}+(-{(1+q^2) \over w})_{,t}=0\\
({(1+p^2) \over w})_{,x}+(-{p\,q \over w})_{,t}=0
\end{eqnarray}

\noindent
These conservation laws are local in the following sense. In general any
conservation law can be written as $X_{,x}=T_{,t}$ , where $X$ and $T$
are functions of h, p,q,r,s,t, and higher derivatives of these functions
with respect $x$ and $t$. Such conservation laws are the local ones.
In the case of nonlocal conservation laws
the functions $X$ and $T$ depend , in addition to h, p,q,r,s,t, and higher
derivatives of these functions with respect $x$ and $t$ , upon the integrals
of these variables with respect to $x$ and $t$. One can find such conservation
laws in this case as well. Let us assume that the function $\Psi$ in (\ref{l1})
-(\ref{l2}) is analytic in the parameter $k$ and can be expanded as

\begin{equation}
\Psi=\Psi_{0}+k\, \Psi_{1}+k^2\, \Psi_{2}+...
\end{equation}

\noindent
then equations  (\ref{l1})-(\ref{l2}) imply

\begin{eqnarray}
\Psi_{0}=g^{-1}\\
(g\, \Psi_{1})_{,x}= - g\,M\,g^{-1}\\
(g\, \Psi_{1})_{,t}= - g\,N\,g^{-1}\\
(g\, \Psi_{2})_{,x}=g_{x}\,g^{-1} - g\,M\,g^{-1}\,D_{x}^{-1}\, g\,M\,g^{-1}\\
(g\, \Psi_{2})_{,t}=g_{t}\,g^{-1} - g\,N\,g^{-1}\,D_{x}^{-1}\, g\,N\,g^{-1}\\
............................................................. \nonumber
\end{eqnarray}

\noindent
where $D_{x}^{-1}$ and $D_{t}^{-1}$ are respectively the inverse operators
of the total derivatives with respect to $x$ and $t$  and

\begin{equation}
M=-r^{\prime}\,g^{-1}g_{,t}+q^{\prime}\,g^{-1}g_{,x}~~~,~~~
N=-q^{\prime}\,g^{-1}g_{t}+p^{\prime}\,g^{-1}g_{x}
\end{equation}

\noindent
Hence we have now infintely many conservation laws with finctions
$X_{n}$ and $T_{n}$ for all $n=0,1,2..$. First two members may
be givem from the above equations

\begin{eqnarray}
X_{0}=M ~~, ~~T_{0}=N\\
X_{1}=g^{-1}\,g_{,x}+(D_{x}^{-1}\,M)\,M ~~,~~T_{1}=g^{-1}\,g_{,t}+
(D_{t}^{-1}\,N)\,N\\
............................................................ \nonumber
\end{eqnarray}

\noindent
In this way one can find infinitely many nonlocal coanservation
laws.


\noindent
{\bf 4.} \,The Backlund transformation obtainable from the Lax pair (\ref{l1})-(\ref{l2})
is not suitable becouse the correspondance between the new and old
solutions will be of the same degree of the degree of the minimality condition.
Hence one has to solve a second order differential equation which
is as hard as the original equation. Instead we shall mention two
interesting nonauto Backlund transformations

The solution of (\ref{met5}) can be expressed interms of two
harmonic functions.

\noindent
{\bf Proposition 3}. Let $x$ and $t$ be harmonic functions of $u$ and $v$
and let a differentiable function $h(t,x)$ be defined by

\begin{eqnarray*}
[1+p^2]\,t,_{u}=-w\,x,_{v}-q\,p\,x,_{u}\\  \nonumber
[1+p^2]\,t,_{v}=-w\,x,_{u}-q\,p\,x,_{v} \label{hod1}
\end{eqnarray*}

\noindent
Then the function $h(t,x)$ is a harmonic function of $u$ and $v$
if and only if it satisfies the minimality condition (\ref{met5}).

\noindent
This proposition implies that the function $h(t,x)$  can be costructed
from (\ref{hod1})
interms of two harmonic functions $t(u,v)$ and $x(u,v)$.
The function $h(t,x)$ obtained this way  satisfies the minimality condition
(\ref{met5}) automatically. In this case the metric (\ref{met6})
on the two dimensional surface $S$ takes the conformally flat form

\begin{equation}
ds_{(m)}^2=w^2\, \bigl({x,_{u}^2+x,_{v}^2 \over 1+p^2} \bigr)\,(du^2+dv^2)
\end{equation}

\noindent
Here we understand that the minimality condition (\ref{met5})
arises from a sigma model so that the target and base space metrics
are the same. Such a sigma model has a Lax pair defined in the
linear equation (\ref{lp1}) in proposition 1 (or in (\ref{l1} - \ref{l2})).
This Lax equation may
be used to construct Backlund transformation for the equation (\ref{met5})
(the minimality condition). Instead of following such a direction we
find the Backlund transformation by defining a new $2 \times 2$ matrix
function Q,

\begin{equation}
g^{\alpha\, \beta}\,g^{-1}\, {\partial_{\beta} \,g}=
\epsilon^{\alpha\, \beta}\, {\partial_{\beta}\, Q}  \label{met7}
\end{equation}

\noindent
{\bf Proposition 4}: (a). Equation corresponding to the matrix $Q$ is

\begin{equation}
\partial_{\alpha}\,(g^{\alpha\, \beta}\, \partial_{\beta}\,Q)-
\epsilon^{\alpha\, \beta}\, \partial_{\alpha}\,Q\,\,\partial_{\beta}\,Q=0.
\label{met8}
\end{equation}

\noindent
(b). The corrsponding linear equation is

\begin{equation}
\epsilon^{\alpha\, \beta}\, \partial_{\beta}\,\, \Psi=
{1 \over k^2+1}\,(k\,\epsilon^{\alpha\, \beta}+g^{\alpha\, \beta})\,
\partial_{\beta}Q\,\, \Psi \label{lp2}
\end{equation}

\noindent
There is a second Backlund transformation
for the Eq.(\ref{met5}) obtainable simply by using either
(\ref{lp1}) or (\ref{lp2}).

\noindent
{\bf Proposition 5}: Let $z=h(t,x)$ define a minimal surface embedded
in the three dimensional Euclidean space $R^{3}$.
The following transformation

\begin{equation}
{h,_{x} \over w}=\psi,_{t}~~~,~~~{h,_{t} \over w}=-\psi,_{x}
\label{bt}
\end{equation}

\noindent
maps the minimality condition (\ref{met5}) to the equation

\begin{equation}
(1-\psi,_{x}^2)\,\psi,_{tt}+2\psi,_{x}\,\psi,_{t}\,\psi,_{xt}+
(1-\psi,_{t}^2)\,\psi,_{xx}=0,
\label{met9}
\end{equation}

\noindent
This equation defines a minimal surface $S^{\prime}=((t,x,w^{\prime}):
w^{\prime}=\psi(t,x))$. $S^{\prime}$ is embedded in a three dimensional
Minkowski space $M_{3}$ with the metric $ds^2=dt^2+dx^2-d\,w^{\prime\,^2}$.
The metric on $S^{\prime}$ is given by

\begin{equation}
ds_{(m)}^{\prime \,2}=g^{\prime}_{(m)\,\alpha \, \beta}\,dx^{\alpha}\,dx^{\beta}\,=
(1-\psi,_{t}^2)\,d\,t^2-2\, \psi,_{t}\, \psi,_{x}\,\,d\,x\,d\,t+
(1-\psi,_{x}^2)\,d\,x^2 \label{met10}
\end{equation}

\noindent
The minimality condition (\ref{met9}) for the surface $S^{\prime}$
may be written as

\begin{equation}
g_{(m)}^{\prime \,\alpha \, \beta}\, \psi,_{\alpha\, \beta}=0.
\end{equation}

\noindent
As an illustration  to the above transformation (\ref{bt})
we can give the following nontrivial examples. The following minimal
surfaces

$$ \displaystyle \psi=
{1 \over \lambda}\,[ln\,cosh(\lambda\,t)-ln\,cosh(\lambda\,x)]$$

\noindent

$$\displaystyle h={1 \over \lambda}\,cos^{-1}\,[sinh(\lambda\,t)\,sinh(\lambda\,x)]$$

\noindent
are transformable to each other. Here $\lambda$ is a nonvanishing constant.

Finally we would like to mention another Backlund transformation wchich
maps solutions of the minimality condition to the solutions of the
Monge-Ampere equation. This is given by the following proposition

\noindent
{\bf Proposition 6:} Let the function $h(t,x)$ with enough differentiability
satisfy the minimality condition (\ref{met5}) then the metric $g_{\mu\, \nu}
={1 \over w}\,g_{(m)\, \mu\, \nu}$ satisfies the condition

\begin{equation}
\partial_{\alpha}\,g_{\mu\, \nu}= \partial_{\nu}\,g_{\mu\, \alpha},
\end{equation}

\noindent
which also implies that

\begin{equation}
g_{\mu\, \nu}= \partial_{\mu}\,\partial_{\nu}\, u , \label{ma1}
\end{equation}

\noindent
where $u(t,x)$ is enough differentiable function of $t,x$ satisfying
the equation

\begin{equation}
Det(\partial_{\mu}\,\partial_{\nu}\, u)= u_{,tt}\,u_{,xx}-u_{tx}^{2}=1.
\end{equation}

\noindent
This is the equation known as the Monge-Amp{\' e}re equation. This equation
is also integrable and its Lax-Pair can be easily obtained by using
(\ref{ma1}) in (\ref{lp1}) or in (\ref{l1}-\ref{l2}). Hyperbolic
minimal surfaces have also similar correspondance with
the Monge-Amp{\' e}re equation. Using (\ref{met9}) and (\ref{met10})
we have

\begin{equation}
g^{\prime}_{\mu\, \nu}= \partial_{\mu}\,\partial_{\nu}\, u , \label{ma2}
\end{equation}

\noindent
with

\begin{equation}
Det(\partial_{\mu}\,\partial_{\nu}\, u)= u_{,tt}\,u_{,xx}-u_{tx}^{2}=1.
\end{equation}

\noindent
which doesnot give the hyperbolic Monge-Amp{\' e}re equation
as expected. The correspondance between the minimal surfaces
in $R^{3}$ and the Monge-Amp{\' e}re equation is mentioned in
\cite{jor}-\cite{hnz}. The correspondance between the Born-infeld
and the hyperbolic Monge-Amp{\' e}re equation is mentioned in
\cite{mok}.

\newpage

\noindent
This work is partially supported by the Scientific and Technical Research
Council of Turkey (TUBITAK) and Turkish Academy of Sciences (TUBA).

\end{document}